\documentstyle[11pt]{article}
\begin{document}

\title{SUSY charged Higgs boson pair production via gluon-gluon collisions}

\author{{
Jiang Yi$^{b}$,
Ma Wen-Gan$^{a,b,c}$, \footnote{e-mail address: mawg@lx04.mphy.ustc.edu.cn}
Han Liang$^{b}$,  Han Meng$^{b}$ and Yu Zeng-Hui$^{b}$}\\
{\small $^{a}$CCAST (World Laboratory), P.O.Box 8730, Beijing 100080,
        P.R.China.}\\
{\small $^{b}$Department of Modern Physics, University of Science
        and Technology of China (USTC)} \\
{\small Hefei, Anhui 230027, P.R.China \footnote{Mailing address}.}\\
{\small $^{c}$Institute of Theoretical Physics, Academia Sinica,}
{\small P.O.Box 2735, Beijing 100080, P.R.China. }
 }
\date{}
\maketitle

\begin{center}\begin{minipage}{100mm}
\vskip 10mm
\begin{center} {\bf Abstract}\end{center}
\baselineskip 0.3in
{The production of charged Higgs boson pairs from gluon-gluon
fusion is studied in the minimal supersymmetric model(MSSM) at
proton-proton colliders. We find that the contribution of the
scalar quark loops to the subprocess production rate is substantial
and consequently the production rate in proton-proton colliders
is enhanced in the MSSM if there exist scalar quarks with favorable
masses. The results show that this cross section may reach a few
femptobarn in the future LHC collider with plausible values of the
parameters.}

\vskip 10mm
{PACS number(s): 14.80.Cp, 12.15.Lk, 12.60.Jv}
\end{minipage}
\end{center}

\newpage
\noindent
{\Large{\bf 1.Introduction}}
\vskip 5mm

\begin{large}
\baselineskip 0.35in
The minimal standard model(MSM) \cite{s1}\cite{s2}, a gauge theory based
on $SU(3)_{c} \times SU(2)_{L} \times U(1)_{Y}$, is a very
successful model which nearly describes either quantitatively
or qualitatively all available data pertaining to the
strong and electroweak interaction phenomena. Until now
the symmetric breaking sector of the electroweak interactions
has not yet been directly tested experimentally. The future multi-TeV hadron
colliders such as the CERN Large Hadron Collider (LHC) and Next Linear
Collider(NLC) are elaborately designed in order to detect a 'light'
scalar boson($m_{H}\leq 1 TeV$) associated with the symmetry-breaking
mechanism.

Any enlargement of the sector beyond the single $SU(2)_{L}$ Higgs doublet
of MSM, with two or more doublets as required in supersymmetric theory,
necessarily involves new physical particles. In this extension theory
the physical spectrum includes charged Higgs bosons. Like the general
two-Higgs-doublet model(THDM), the minimal supersymmetric standard
model(MSSM) also requires two $SU(2)$ doublets to give masses to up
and down quarks \cite{s3}\cite{haber}. Since both models contain the
same Higgs sector structure, it is of interest to determine whether
THDM or MSSM Higgs sector might appear in future experiments.

 With regard to colliders in the future there are several mechanisms which can
produce charged Higgs bosons. Among them the charged Higgs boson pair
productions via gluon-gluon collision, quark-antiquark annihilation
and $\gamma \gamma$ collision, play a significant role in studying
charged Higgs bosons. Similar to the neutral Higgs pair production
process in hadron colliders, the charged Higgs pair production process
also involves Higgs sector interactions, therefore the measurement
of these kinds of couplings could be a good test of our theory. Recently
Plehn $et~al$ provided calculations concerning the pair
production of neutral Higgs particles in gluon-gluon collisions in
MSSM\cite{s4} and charged Higgs boson pair production at NLC in
$\gamma \gamma$ modes are also investigated in references \cite{s5}.
The charged Higgs boson pair production via gluon-gluon fusion in
general THDM is also represented in Ref.\cite{s6}. In this reference, it
is concluded that the production rate of this process in THDM is not
sensitive to the neutral Higgs boson masses and the couplings of
self interactions of Higgs bosons, but it is strongly related to
the couplings of the charged Higgs boson to quarks.
Willenbrock discussed the charged Higgs boson pair production
via gluon fusion in MSSM in Ref.\cite{s8}. He showed that the gluon fusion
via quark and scalar-quark loops would dominate the usual electroweak
quark pair annihilation process if there exist sufficiently heavy quarks
or scalar-quarks. In his work, he took some approximations, such as the
large quark and squark mass limits, i.e., $m_{q}^2 (m_{\tilde{q}}^2)
>> \hat{s}$, in numerical calculations. After we have finished our 
paper, we found a work also on this process. One can refer to
Ref.\cite{Krause}.

In this paper we concentrate on the capability of the charged Higgs
boson pair production via gluon-gluon collisions at $pp$ colliders
in frame of MSSM without the approximations used in reference \cite{s8}.
The paper is organized as follows: The analytical
evaluations are given in section 2. In section 3 there are numerical
results, discussion and a short summary. Finally the explicit
expressions of the relevant form factors induced by the supersymmetric
quark loop diagrams are collected in Appendix.

\vskip 5mm
\noindent
{\Large{\bf 2.Calculation}}
\vskip 5mm
The MSSM theory gives a large number of parameters, since it predicts
more than doubling the MSM spectrum of particle states. The number of the
parameters can be reduced largely by embedding the low-energy
supersymmetric theory into a grand unified(GUT) framework. Besides
the parameters which can be achieved in supergravity model\cite{s7},
we need extra parameters to describe the Higgs sector: one Higgs mass
parameter and the two vacuum expectation values of the neutral Higgs
fields. We can also use $\tan{\beta}=\frac{v_2}{v_1}$, the ratio of
the vacuum expectation values of the two-Higgs-doublet fields which
break the electroweak symmetry. If we ignore the flavor mixing,
the mass matrix of a scalar quark takes the following form\cite{s9}:

\begin{equation}
\label{e1}
 -{\cal L}_{m}=\left( \begin{array}{ll}
                        \tilde{q}^{\ast}_{L} &  \tilde{q}^{\ast}_{R}
                    \end{array} \right)
       \left( \begin{array}{ll}  m^2_{\tilde{q}_{L}} &  a_{q} m_{q} \\
                           a_{q}^{\ast} m_{q}  &  m^2_{\tilde{q}_{R}}
              \end{array}  \right)
       \left( \begin{array}{ll}  \tilde{q}_{L}  \\  \tilde{q}_{R}
              \end{array}  \right)
\end{equation}

here $\tilde{q}_{L}$ and $\tilde{q}_{R}$ are the current eigenstates and
for the up-type scalar quarks,
$$m^2_{\tilde{q}_{L}}=\tilde{M}^2_{Q} + m_{Z}^2 \cos{2 \beta}(\frac{1}{2}-
      Q_q \sin{\theta_W}^2) + m^2_{q}, $$
$$m^2_{\tilde{q}_{R}}=\tilde{M}^2_{U} + m_{Z}^2 \cos{2 \beta} Q_q
      \sin{\theta_W}^2 + m^2_{q},  $$
\begin{equation}
\label{e2}
a_{q}=\mu \cot{\beta} + A^{\ast}_{q} \tilde{M},
\end{equation}
for the down-type scalar quarks,
$$m^2_{\tilde{q}_{L}}=\tilde{M}^2_{Q} - m_{Z}^2 \cos{2 \beta}(\frac{1}{2}+
      Q_q \sin{\theta_W}^2) + m^2_{q}, $$
$$m^2_{\tilde{q}_{R}}=\tilde{M}^2_{D} + m_{Z}^2 \cos{2 \beta} Q_q
      \sin{\theta_W}^2 + m^2_{q},  $$
\begin{equation}
\label{e3}
a_{q}=\mu \tan{\beta} + A^{\ast}_{q} \tilde{M},
\end{equation}

where $Q_{q}$ is the charge of the scalar quark, $\tilde{M}^2_{Q}$,
$\tilde{M}^2_{U}$ and $\tilde{M}^2_{D}$ are the self-supersymmetry-breaking
mass terms for the left-handed and right-handed scalar quarks.
$A_{q}\times\tilde{M}$ is a trilinear scalar interaction parameter,
and $\mu$ is the supersymmetric mass mixing term of the Higgs boson.
When there is no CP violation, we can regard that $a_{q}$ is a real constant.
At the Plank mass scale we take

\begin{equation}
\label{e4}
\tilde{M}^2_{Q}=\tilde{M}^2_{U}=\tilde{M}^2_{D}=\tilde{M}^2.
\end{equation}

The mass eigenstates $\tilde{q}_1$ and $\tilde{q}_2$ are related to the current
eigenstates $\tilde{q}_L$ and $\tilde{q}_R$  by
$$
\tilde{q}_1=\tilde{q}_L \cos{\theta_q} + \tilde{q}_R \sin{\theta_q}
$$
\begin{equation}
\label{e5}
\tilde{q}_2=- \tilde{q}_L \sin{\theta_q} + \tilde{q}_R \cos{\theta_q}.
\end{equation}

and

\begin{equation}
\label{e6}
\tan{2 \theta_q}=\frac{2 a_{q} m_{q}} {m^2_{\tilde{q}_{R}}-m^2_{\tilde{q}_{L}} }.
\end{equation}

The masses of $\tilde{q}_1$ and $\tilde{q}_2$ are

\begin{equation}
\label{e7}
(m^2_{\tilde{q}_1},m^2_{\tilde{q}_2})=\frac{1}{2} {
     m^2_{\tilde{q}_L} + m^2_{\tilde{q}_R} \mp  [ (m^2_{\tilde{q}_L} -
     m^2_{\tilde{q}_R})^2 + 4 a_q^2 m_q^2 ] ^{\frac{1}{2}} }.
\end{equation}

 In the evaluation we neglected the first SUSY generation mixing angle,
whereas in the second SUSY generation we took $\sin{\theta_q} = 0.2$,
and let the third SUSY generation has the mixing with
$\sin{\theta_q} \approx \frac{1}{\sqrt{2}}$ due to the large splitting
of the mass eigenstates. The numerical value of the lightest
physical stop quark mass is taken as $m_{\tilde{t}_1}=180 GeV$ and
the sbottom quark mass as $m_{\tilde{b}_1}=400 GeV$. We set the masses
of the squarks of $\tilde{u}_1$, $\tilde{d}_1$, $\tilde{c}_1$ and
$\tilde{s}_1$ degenerated with mass value $800 GeV$. As a
quantitative example, we took $\mu=0$ in our calculation.

The generic Feynman diagrams contributing to the subprocess $gg \rightarrow
H^{+} H^{-}$ which are involved in frame of MSSM are depicted in figure 1.
The diagrams created by exchanging the two external gluon lines are not shown in Fig.1.
Fig.1$(1 \sim 10)$ are the same as those in the THDM model\cite{s6}, we call
them as "THDM part". Fig.1$(11 \sim 32)$ are the diagrams with squark-loop.
There is no tree level contribution for this subprocess, therefore the proper
vertex counterterm cancels with the counterterms of the external legs diagrams.
That is to say that the evolution can be simply carry out by summing all
unrenormalized reducible and irreducible diagrams. The result is finite and
gauge invariant. The relevant Feynman rules can be
found in references\cite{s9}\cite{s10}. Since the main contributions are
due to the loops of heavy quarks and the scalar quarks, we neglect the
light quark loop diagrams. We can see from Fig.1 that there are mainly
three kinds of production mechanisms. One  is by exchanging a
photon or a $Z^0$ boson to produce a pair of charged Higgs bosons, we
refer to this as $\gamma-Z^0$ exchange s-channel (see Fig.1(1,2), Fig.1(11,12),
Fig.1(15,16) and Fig.1(19,20)). For the $\gamma-Z^0$ exchange s-channel
diagrams, the contributions are shown to be zero.
This is the consequence of Furry theorem. Since the Furry theorem forbids
the production of the spin-one components of the $Z^{0}$ and $\gamma$,
and the spin-zero component of the $Z^{0}$ vector boson does not couple to
a pair of scalar particles with the same mass. The second mechanism is
through producing the virtual neutral Higgs $h^0$ and $H^0$ bosons which are
coupled to gluons by quartic coupling including squarks or by a
triangle of heavy quarks or squarks, then the charged Higgs boson
pair is appeared by the subsequent decay of the virtual Higgs bosons
(see Fig.1(3,4,13,14,17,18)). It is called neutral Higgs boson exchange
s-channel. An alternative is that the charged Higgs bosons are
produced by means of the box and quartic interactions.

In this paper we follow the notations in reference \cite{s6} that
$p_{1}$, $p_{2}$, $p_{3}$ and $p_{4}$ represent the four-momenta
of the final charged Higgs bosons and the initial gluons, respectively.
We write the corresponding matrix element for each of the diagrams
in Fig.1 in the form as
\begin{eqnarray*}
M_{l} &=& \frac{-1}{16 \pi^2} g_{s}^{2} \epsilon_{\mu}(p_3)
           \epsilon_{\nu}(p_4) A_{2} \cdot \\
    & &  (f_{1,l} g^{\mu \nu} + f_{2,l} p_1^{\mu} p_1^{\nu} +
          f_{3,l} p_1^{\mu} p_2^{\nu} +
          f_{4,l} p_2^{\mu} p_1^{\nu} + f_{5,l} p_2^{\mu} p_2^{\nu} + \\
    & &   f_{6,l} \epsilon^{\mu \nu \alpha \beta} p_{1 \alpha} p_{2 \beta} +
          f_{7,l} \epsilon^{\mu \nu \alpha \beta} p_{1 \alpha} p_{3 \beta} +
          f_{8,l} \epsilon^{\mu \nu \alpha \beta} p_{2 \alpha} p_{3 \beta} + \\
    & &   f_{9,l} \epsilon^{\nu \alpha \beta \gamma} p_{1 \alpha}
                    p_{2 \beta} p_{3 \gamma} p_1^{\mu} +
          f_{10,l} \epsilon^{\mu \alpha \beta \gamma} p_{1 \alpha}
                    p_{2 \beta} p_{3 \gamma} p_1^{\nu} +
\end{eqnarray*}
$$
   f_{11,l} \epsilon^{\nu \alpha \beta \gamma} p_{1,\alpha}
   p_{2,\beta} p_{3,\gamma} p_2^{\mu} +
   f_{12,l} \epsilon^{\mu \alpha \beta \gamma} p_{1,\alpha}
   p_{2,\beta} p_{3,\gamma} p_2^{\nu} ).   \eqno(8)
$$
We denote $A_{k}$(k=1,2) as the color factors of the diagrams with and
without gluon-gluon-squark-squark quartic vertex respectively and it can be
expressed as

\begin{eqnarray*}
A_{1} &=&  Tr[\frac{1}{3} \delta_{ab} I + d_{abc} T^{c}] = \delta_{ab},
\end{eqnarray*}
$$
A_{2} = Tr[T^a T^b]=\frac{1}{2}\delta_{ab}. \eqno(9)
$$

In above equations $I$ is a $3 \times 3$ unit matrix.
$g_s$ is the strong coupling constant and $T^a(a=1 \sim 8)$ are the
$SU(3)_c$ generators introduced by Gell-Mann and $d_{abc}$ are the
structure constants\cite{s11}. Here we should declare that in Eq.(8)
$A_{2}=\frac{1}{2}\delta_{ab}$ is the common factor of the all color
factors in amplitudes and the remained parts of color factors are
collected in form factors.

The total matrix element of subprocess $gg \rightarrow H^{+} H^{-}$
in MSSM can be expressed as
$$
\begin{array}{lll}
M &=& \sum\limits_{l=1}^{10} M_{l} + \sum\limits_{n=11}^{32} M_{n} \\
  &=& M_{THDM} + M_{SUSY},
\end{array}
  \eqno(10)
$$
where $M_{THDM}$ represents the summation of the amplitudes of the THDM part
and $M_{SUSY}$ is the total amplitudes of squark-loop diagrams.

Then we can split each form factor appeared in total matrix element up
into two parts.
$$
f_{i}=f_{i}^{THDM}+ f_{i}^{SUSY}, (i=1 \sim 12).  \eqno(11)
$$
where $f_{i}^{SUSY}(i=1 \sim 12)=\sum\limits_{n=11}^{32} f_{i,n}$
and $f_{i}^{THDM}(i=1 \sim 12)= \sum\limits_{l=1} ^{10} f_{i,l}$.
The explicit expressions of the form factors $f_{i}^{SUSY}(i=1 \sim 12)$,
which come from the squark-loop diagrams, are listed in Appendix.
For the form factors of the THDM part(i.e. $f_{i}^{THDM}(i=1 \sim 12)$)
one can find them in reference\cite{s6}. The total cross section
for subprocess $gg \rightarrow H^{+}H^{-}$ can finally be written in
the form
$$
\hat{\sigma}(\hat{s})=\frac{1}{16 \pi \hat{s}^2}
\int_{\hat{t}^{-}}^ {\hat{t}^{+}} d\hat{t} \bar{\vert M \vert}^2. \eqno(12)
$$
where $\bar{\vert M \vert}^2$ is the initial spin and color averaged matrix
element squared and $\hat{t}^{\pm}=(m_{H^{\pm}}^2-\frac{1}{2}\hat{s})
\pm \frac{1}{2}\hat{s} \beta_{H^{\pm}}$.
The total cross section for the charged Higgs pair production
through gluon-gluon fusion in proton-proton collisions can be
obtained by integrating the $\hat{\sigma}$ over the gluon luminosity.
$$
\sigma(pp \rightarrow gg \rightarrow H^{+}H^{-}+X)=
    \int_{4 m_{H^{\pm}}^2/s }^{1}
d\tau \frac{dL_{gg}}{d\tau} \hat{\sigma}(gg \rightarrow
H^{+}H^{-} \hskip 3mm at \hskip 3mm \hat{s}=\tau s). \eqno(13)
$$

where $\sqrt{s}$ and $\sqrt{\hat{s}}$  denote the proton-proton
and gluon-gluon c.m. energies respectively and
$\frac{dL_{gg}} {d\tau}$ is the gluon luminosity, which is defined as

$$
\frac{d{\cal L}_{gg}}{d\tau}=\int_{\tau}^{1}
 \frac{dx_1}{x_1} \left[ F_{g}(x_1,\mu) F_{g}(\frac{\tau}{x_1},\mu) \right].
 \eqno(14)
$$
Here we used $\tau=x_{1}x_{2}$, and one can find the definitions of
$x_1$ and $x_2$ in Ref.\cite{s6}.
In our calculation we adopt the MRS set G parton distribution function
$F_{g}(x)$ \cite{ss1}, and ignore the supersymmetric QCD corrections
to the parton distribution functions for simplicity. The numerical calculation
is carried out for the LHC at the energy around $10 \sim 14 TeV$.

\vskip 5mm
\noindent
{\Large{\bf 3. Numerical results and discussion}}
\vskip 5mm

For the numerical evaluation we take the input parameters\cite{s12} as
$m_b=4.5~ GeV,~m_Z=91.1887~GeV,~m_W=80.2226~GeV$,$m_t=175~GeV$ ,
$G_F=1.166392\times 10^{-5}(GeV)^{-2} ~and ~\alpha=\frac{1}{137.036}$.
We adopt a simple one-loop formula for the running strong coupling
constant $\alpha_s$ as
$$
\alpha_s(\mu)=\frac{\alpha_{s}(m_Z)} {1+\frac{33-2 n_f} {6 \pi} \alpha_{s}
              (m_Z) \ln \left( \frac{\mu}{m_Z} \right) }. \eqno(15)
$$
where $\alpha_s(m_Z)=0.117$ and $n_f$ is the number of active flavors at
energy scale $\mu$. The two representative quantities of the two vacuum
expectation values $\tan{\beta}=1.5$ and $\tan{\beta}=30$ are chosen.

We take the mass of charged Higgs bosons $m_{H^{\pm}}$, and the ratio of
vacuum expectation values as free parameters in MSSM.  The other Higgs boson
masses and the mixing angle $\alpha$ are fixed by quoting the MSSM
relations in terms of $\tan{\beta}$ and $m_{H^{\pm}}$. The relevant formulas
are given as

$$
\tan{2 \alpha}=
    \tan{2 \beta}
    \frac{m_A^2 + m_Z^2}{m_A^2- m_Z^2},~~ where~~ m_A^2 = m_{H^{\pm}}^2 - m_W^2.
    \eqno(16)
$$
$$
m^2_{H^0,h^0}=\frac{1}{2} \left[ m_{A}^2+m_{Z}^2 \pm \sqrt{(m_A^2+m_Z^2)^2-
        4 m_Z^2 m_A^2 cos^2 2 \beta } \right].   \eqno(17)
$$

The cross sections of the subprocess $gg \rightarrow H^{+} H^{-}$
with $m_{H^{\pm}}=150~GeV$  are depicted in Fig.2(1), on the conditions
of the mass values of squarks as mentioned above. In this figure,
the small peak around $\sqrt{\hat{s}} \sim 350~GeV$ of each curve
originates from the resonance effect, where $\sqrt{\hat{s}} \sim
2 m_{t} $ and $\sqrt{\hat{s}} \sim  2 m_{\tilde{t}_1} $ are satisfied.
Around $\sqrt{\hat{s}} \sim 800~GeV$,
there is another higher peak for each curve, which comes from the enhancement
of the resonance effects where the c.m. energy of gluon-gluon system is
located at $\sqrt{\hat{s}} \sim 2 m_{\tilde{t}_2}  \sim 800~GeV$ and
$\sqrt{\hat{s}} \sim 2 m_{\tilde{b}_1} \sim 800~GeV$.
Fig.2(2) shows the cross section of this subprocess as the functions
of c.m. energy of incoming gluons with the same neutral Higgs boson, quark
and squark masses as in Fig.2(1), but $m_{H^{\pm}}=300~GeV$. In Fig.2(2)
again the resonance effects of the squarks are obvious at the position near
$\sqrt{\hat{s}} \sim 800~GeV$.

  The dependence of the subprocess cross section to the ratio of vacuum
expectation values, is plotted in Fig.3 with $m_{H^{+}}=150 GeV$. The
two curves correspond to $\sqrt{\hat{s}}=400 GeV$ and $\sqrt{\hat{s}}=800 GeV$,
respectively. They all have the minimal cross sections when $\tan{\beta}$
is around 6.5. The figure shows that the cross section of the
subprocess increases when the $\sqrt{\hat{s}}$ goes up from $400~GeV$ to
$800~GeV$. The dependence of the subprocess cross section to the mass of
charged Higgs bosons is shown in Fig.4 with $\sqrt{\hat{s}}=1 TeV$.
The cross section increases in the region where the charged Higgs
boson mass has the value approximately larger than $\sim 300~GeV$.
That is because when the value of the the charged Higgs boson mass
approaches $\frac{1}{2} \sqrt{\hat{s}}=500~GeV$, the production rate will
be enhanced by the threshold effect, but when the mass of charged Higgs
boson is very close to $500 GeV$, the cross section drops down quickly due to
the phase space suppression.
The cross sections of process $pp \rightarrow gg \rightarrow H^{+} H^{-}+X$
in the LHC energy regions in both MSSM and THDM are depicted in
Fig.5(1$\sim$2). The charged Higgs boson pair production rates in MSSM
are read to be about 1.9 to 5.2 femptobarn at the LHC energy
regions when $m_{H^{+}}=150 GeV$. With the heavier charged Higgs mass, the
production rate will be smaller. The production rates in the MSSM are
higher than those in the THDM with the same parameters.
In our calculation, we also find that among the
squark-loop diagrams, the third generation squarks make the decisive
contribution, while the contribution from the other squarks is rather
small comparatively due to the decoupling of the heavy squarks.

In conclusion, we investigated the pair production process of charged
Higgs bosons via gluon-gluon fusion in pp collider at LHC.
The numerical analysis of its cross section is carried out in the MSSM
model. With the possible parameters, the cross sections at future
LHC collider may reach 5.2 femptobarn. The calculation shows that
the contribution from the squark-loop diagrams is obvious
and consequently enhances the cross section of $pp \rightarrow
gg \rightarrow H^{+} H^{-} + X$, if there exist scalar quarks with
favorable masses.

\par We express our thanks to Dr. M. Spira for useful discussions. This
work was supported by the National Natural Science Foundation of China.
Part of this work was done when two of the authors, Ma Wen-Gan and Yu
Zeng-Hui, visited the University Vienna under an exchange
agreement(project number: IV.B.12). 

\vskip 5mm
\noindent
{\Large{\bf Appendix}}
\vskip 5mm

The form factors involved in equation (8) are presented explicitly as:

\begin{eqnarray*}
f_{1}&=& -2g^{2}_{s} \sum\limits_{i=1,2}\{
  \sum\limits_{\tilde{q}_{i}=\tilde{u}_{i},\tilde{d}_{i},
  \tilde{c}_{i}}^{\tilde{s}_{i}, \tilde{t}_{i},\tilde{b}_{i}}
(g_{h^{0}H^{\pm}H^{\mp}}g_{h^{0}\tilde{q}_{i}\tilde{q}_{i}}A_{h^{0}} +
    g_{H^{0}H^{\pm}H^{\mp}}g_{H^{0}\tilde{q}_{i}\tilde{q}_{i}}A_{H^{0}}
+ g_{H^{+}H^{-}\tilde{q}_{i}\tilde{q}_{i}})\\
&&\{4 C_{24}[-p_3, p_1 + p_2, m_{\tilde{q}_{i}}, m_{\tilde{q}_{i}}, m_{\tilde{q}_{i}}]
- B_{0}[-p_1 - p_2, m_{\tilde{q}_{i}}, m_{\tilde{q}_{i}}] \}\\
&+&
2\sum\limits_{j=1,2} \sum\limits_{(\tilde{q}_{u,i},\tilde{q}_{d,j})=
(\tilde{u}_{i},\tilde{d}_{j})}
  ^{(\tilde{c}_{i},\tilde{s}_{j}),(\tilde{t}_{i},\tilde{b}_{j})}
g^{2}_{H^{\pm}\tilde{q}_{u,i}\tilde{q}_{d,j}} \{
-\frac{1}{2}C_{0}[p_{1}+p_{2}, -p_{2}, m_{\tilde{q}_{u,i}}, m_{\tilde{q}_{u,i}},
 m_{\tilde{q}_{d,i}}]\\
&-&\frac{1}{2}C_{0}[p_{1}+p_{2}, -p_{2}, m_{\tilde{q}_{d,i}}, m_{\tilde{q}_{d,i}},
 m_{\tilde{q}_{u,i}}]\\
&+&i D_{27}[-p_3, p_1 + p_2, -p_2,
m_{\tilde{q}_{u,i}}, m_{\tilde{q}_{u,i}}, m_{\tilde{q}_{u,i}}, m_{\tilde{q}_{d,j}}]\\
&+&i D_{27}[-p_3, p_1 + p_2, -p_2,
m_{\tilde{q}_{d,j}}, m_{\tilde{q}_{d,j}}, m_{\tilde{q}_{d,j}}, m_{\tilde{q}_{u,i}}]\\
&+&i D_{27}[p_3, -p_1 - p_2, p_1,
m_{\tilde{q}_{u,i}}, m_{\tilde{q}_{u,i}}, m_{\tilde{q}_{u,i}}, m_{\tilde{q}_{d,j}}]\\
&+&i D_{27}[p_3, -p_1 - p_2, p_1,
m_{\tilde{q}_{d,j}}, m_{\tilde{q}_{d,j}}, m_{\tilde{q}_{d,j}}, m_{\tilde{q}_{u,i}}]\\
&+& D_{27}[p_1, p_2 - p_3, -p_2,
m_{\tilde{q}_{d,j}}, m_{\tilde{q}_{u,i}}, m_{\tilde{q}_{d,j}}, m_{\tilde{q}_{u,i}}]\\
&+& D_{27}[p_1, p_2 - p_3, -p_2,
m_{\tilde{q}_{u,i}}, m_{\tilde{q}_{d,j}}, m_{\tilde{q}_{u,i}}, m_{\tilde{q}_{d,j}}]\}\}
\hskip 33mm (A-1)
\end{eqnarray*}

\begin{eqnarray*}
f_{2}&=& -2g^{2}_{s} \sum\limits_{i=1,2}\{
  \sum\limits_{\tilde{q}_{i}=\tilde{u}_{i},\tilde{d}_{i},
  \tilde{c}_{i}}^{\tilde{s}_{i}, \tilde{t}_{i},\tilde{b}_{i}}
(g_{h^{0}H^{\pm}H^{\mp}}g_{h^{0}\tilde{q}_{i}\tilde{q}_{i}}A_{h^{0}} +
    g_{H^{0}H^{\pm}H^{\mp}}g_{H^{0}\tilde{q}_{i}\tilde{q}_{i}}A_{H^{0}}
+ g_{H^{+}H^{-}\tilde{q}_{i}\tilde{q}_{i}})\\
&& (C_{23}-C_{22})[-p_3, p_1 + p_2, m_{\tilde{q}_{i}},
m_{\tilde{q}_{i}}, m_{\tilde{q}_{i}}] \\
&+&
\sum\limits_{j=1,2} \sum\limits_{(\tilde{q}_{u,i},\tilde{q}_{d,j})=
(\tilde{u}_{i},\tilde{d}_{j})}
  ^{(\tilde{c}_{i},\tilde{s}_{j}),(\tilde{t}_{i},\tilde{b}_{j})}
g^{2}_{H^{\pm}\tilde{q}_{u,i}\tilde{q}_{d,j}} \{
2i (D_{24}\\
&&-D_{22})[-p_3, p_1 + p_2, -p_2,
m_{\tilde{q}_{u,i}}, m_{\tilde{q}_{u,i}}, m_{\tilde{q}_{u,i}}, m_{\tilde{q}_{d,j}}]\\
&+&2i (D_{24}-D_{22})[-p_3, p_1 + p_2, -p_2,
m_{\tilde{q}_{d,j}}, m_{\tilde{q}_{d,j}}, m_{\tilde{q}_{d,j}}, m_{\tilde{q}_{u,i}}]\\
&+&2i (D_{24}+2D_{26}-D_{22}-D_{23}-D_{25})  [p_3, -p_1 - p_2, p_1,
m_{\tilde{q}_{u,i}}, m_{\tilde{q}_{u,i}}, m_{\tilde{q}_{u,i}}, m_{\tilde{q}_{d,j}}]\\
&+&2i (D_{24}+2D_{26}-D_{22}-D_{23}-D_{25}) [p_3, -p_1 - p_2, p_1,
m_{\tilde{q}_{d,j}}, m_{\tilde{q}_{d,j}}, m_{\tilde{q}_{d,j}}, m_{\tilde{q}_{u,i}}]\\
&-& (D_{0}+3D_{11}+2D_{21})[p_1, p_2 - p_3, -p_2,
m_{\tilde{q}_{d,j}}, m_{\tilde{q}_{u,i}}, m_{\tilde{q}_{d,j}}, m_{\tilde{q}_{u,i}}]\\
&-& (D_{0}+3D_{11}+2D_{21})[p_1, p_2 - p_3, -p_2,
m_{\tilde{q}_{u,i}}, m_{\tilde{q}_{d,j}}, m_{\tilde{q}_{u,i}}, m_{\tilde{q}_{d,j}}]\}\}
\hskip 15mm (A-2)
\end{eqnarray*}

\begin{eqnarray*}
f_{3}&=& -2g^{2}_{s} \sum\limits_{i=1,2}\{
  \sum\limits_{\tilde{q}_{i}=\tilde{u}_{i},\tilde{d}_{i},
  \tilde{c}_{i}}^{\tilde{s}_{i}, \tilde{t}_{i},\tilde{b}_{i}}
(g_{h^{0}H^{\pm}H^{\mp}}g_{h^{0}\tilde{q}_{i}\tilde{q}_{i}}A_{h^{0}} +
    g_{H^{0}H^{\pm}H^{\mp}}g_{H^{0}\tilde{q}_{i}\tilde{q}_{i}}A_{H^{0}}
+ g_{H^{+}H^{-}\tilde{q}_{i}\tilde{q}_{i}})\\
&& (C_{23}-C_{22})[-p_3, p_1 + p_2, m_{\tilde{q}_{i}},
m_{\tilde{q}_{i}}, m_{\tilde{q}_{i}}] \\
&+&
\sum\limits_{j=1,2} \sum\limits_{(\tilde{q}_{u,i},\tilde{q}_{d,j})=
(\tilde{u}_{i},\tilde{d}_{j})}
  ^{(\tilde{c}_{i},\tilde{s}_{j}),(\tilde{t}_{i},\tilde{b}_{j})}
g^{2}_{H^{\pm}\tilde{q}_{u,i}\tilde{q}_{d,j}} \{
2i (D_{24}+D_{26}\\
&&-D_{22})[-p_3, p_1 + p_2, -p_2,
m_{\tilde{q}_{u,i}}, m_{\tilde{q}_{u,i}}, m_{\tilde{q}_{u,i}}, m_{\tilde{q}_{d,j}}]\\
&+&2i (D_{24}+D_{26}-D_{22})[-p_3, p_1 + p_2, -p_2,
m_{\tilde{q}_{d,j}}, m_{\tilde{q}_{d,j}}, m_{\tilde{q}_{d,j}}, m_{\tilde{q}_{u,i}}]\\
&+&2i (D_{24}+D_{26}-D_{22}-D_{25}) [p_3, -p_1 - p_2, p_1,
m_{\tilde{q}_{u,i}}, m_{\tilde{q}_{u,i}}, m_{\tilde{q}_{u,i}}, m_{\tilde{q}_{d,j}}]\\
&+&2i (D_{24}+D_{26}-D_{22}-D_{25}) [p_3, -p_1 - p_2, p_1,
m_{\tilde{q}_{d,j}}, m_{\tilde{q}_{d,j}}, m_{\tilde{q}_{d,j}}, m_{\tilde{q}_{u,i}}]\\
&-&(D_{0}+D_{11}+2D_{12}-2D_{13}+2D_{24} -2D_{25}) [p_1, p_2 - p_3, -p_2,
m_{\tilde{q}_{d,j}}, m_{\tilde{q}_{u,i}}, m_{\tilde{q}_{d,j}}, m_{\tilde{q}_{u,i}}]\\
&-& (D_{0}+D_{11}+2D_{12}-2D_{13}+2D_{24} -2D_{25})[p_1, p_2 - p_3, -p_2,
m_{\tilde{q}_{u,i}}, m_{\tilde{q}_{d,j}}, m_{\tilde{q}_{u,i}}, m_{\tilde{q}_{d,j}}]\}\}
\hskip 5mm (A-3)
\end{eqnarray*}

\begin{eqnarray*}
f_{4}&=& -2g^{2}_{s} \sum\limits_{i=1,2}\{
  \sum\limits_{\tilde{q}_{i}=\tilde{u}_{i},\tilde{d}_{i},
  \tilde{c}_{i}}^{\tilde{s}_{i}, \tilde{t}_{i},\tilde{b}_{i}}
(g_{h^{0}H^{\pm}H^{\mp}}g_{h^{0}\tilde{q}_{i}\tilde{q}_{i}}A_{h^{0}} +
    g_{H^{0}H^{\pm}H^{\mp}}g_{H^{0}\tilde{q}_{i}\tilde{q}_{i}}A_{H^{0}}
+ g_{H^{+}H^{-}\tilde{q}_{i}\tilde{q}_{i}})\\
&& (C_{23}-C_{22})[-p_3, p_1 + p_2, m_{\tilde{q}_{i}},
m_{\tilde{q}_{i}}, m_{\tilde{q}_{i}}] \\
&+&
\sum\limits_{j=1,2} \sum\limits_{(\tilde{q}_{u,i},\tilde{q}_{d,j})=
(\tilde{u}_{i},\tilde{d}_{j})}
  ^{(\tilde{c}_{i},\tilde{s}_{j}),(\tilde{t}_{i},\tilde{b}_{j})}
g^{2}_{H^{\pm}\tilde{q}_{u,i}\tilde{q}_{d,j}} \{
2i (D_{24}+D_{26}\\
&&-D_{22})[-p_3, p_1 + p_2, -p_2,
m_{\tilde{q}_{u,i}}, m_{\tilde{q}_{u,i}}, m_{\tilde{q}_{u,i}}, m_{\tilde{q}_{d,j}}]\\
&+&2i (D_{24}+D_{26}-D_{22})[-p_3, p_1 + p_2, -p_2,
m_{\tilde{q}_{d,j}}, m_{\tilde{q}_{d,j}}, m_{\tilde{q}_{d,j}}, m_{\tilde{q}_{u,i}}]\\
&+&2i (D_{24}+D_{26}-D_{22}-D_{25}) [p_3, -p_1 - p_2, p_1,
m_{\tilde{q}_{u,i}}, m_{\tilde{q}_{u,i}}, m_{\tilde{q}_{u,i}}, m_{\tilde{q}_{d,j}}]\\
&+&2i (D_{24}+D_{26}-D_{22}-D_{25}) [p_3, -p_1 - p_2, p_1,
m_{\tilde{q}_{d,j}}, m_{\tilde{q}_{d,j}}, m_{\tilde{q}_{d,j}}, m_{\tilde{q}_{u,i}}]\\
&-&(D_{12}-D_{13}+2D_{24} -2D_{25}) [p_1, p_2 - p_3, -p_2,
m_{\tilde{q}_{d,j}}, m_{\tilde{q}_{u,i}}, m_{\tilde{q}_{d,j}}, m_{\tilde{q}_{u,i}}]\\
&-&(D_{12}-D_{13}+2D_{24} -2D_{25})[p_1, p_2 - p_3, -p_2,
m_{\tilde{q}_{u,i}}, m_{\tilde{q}_{d,j}}, m_{\tilde{q}_{u,i}}, m_{\tilde{q}_{d,j}}]\}\}
\hskip 15mm (A-4)
\end{eqnarray*}

\begin{eqnarray*}
f_{5}&=& -2g^{2}_{s} \sum\limits_{i=1,2}\{
  \sum\limits_{\tilde{q}_{i}=\tilde{u}_{i},\tilde{d}_{i},
  \tilde{c}_{i}}^{\tilde{s}_{i}, \tilde{t}_{i},\tilde{b}_{i}}
(g_{h^{0}H^{\pm}H^{\mp}}g_{h^{0}\tilde{q}_{i}\tilde{q}_{i}}A_{h^{0}} +
    g_{H^{0}H^{\pm}H^{\mp}}g_{H^{0}\tilde{q}_{i}\tilde{q}_{i}}A_{H^{0}}
+ g_{H^{+}H^{-}\tilde{q}_{i}\tilde{q}_{i}})\\
&& (C_{23}-C_{22})[-p_3, p_1 + p_2, m_{\tilde{q}_{i}},
m_{\tilde{q}_{i}}, m_{\tilde{q}_{i}}] \\
&+&
\sum\limits_{j=1,2} \sum\limits_{(\tilde{q}_{u,i},\tilde{q}_{d,j})=
(\tilde{u}_{i},\tilde{d}_{j})}
  ^{(\tilde{c}_{i},\tilde{s}_{j}),(\tilde{t}_{i},\tilde{b}_{j})}
g^{2}_{H^{\pm}\tilde{q}_{u,i}\tilde{q}_{d,j}} \{
2i (D_{24}\\
&&-D_{22})[-p_3, p_1 + p_2, -p_2,
m_{\tilde{q}_{u,i}}, m_{\tilde{q}_{u,i}}, m_{\tilde{q}_{u,i}}, m_{\tilde{q}_{d,j}}]\\
&+&2i (D_{24}-D_{22})[-p_3, p_1 + p_2, -p_2,
m_{\tilde{q}_{d,j}}, m_{\tilde{q}_{d,j}}, m_{\tilde{q}_{d,j}}, m_{\tilde{q}_{u,i}}]\\
&+&2i (D_{24}+2D_{26}-D_{22}-D_{23}-D_{25})  [p_3, -p_1 - p_2, p_1,
m_{\tilde{q}_{u,i}}, m_{\tilde{q}_{u,i}}, m_{\tilde{q}_{u,i}}, m_{\tilde{q}_{d,j}}]\\
&+&2i (D_{24}+2D_{26}-D_{22}-D_{23}-D_{25}) [p_3, -p_1 - p_2, p_1,
m_{\tilde{q}_{d,j}}, m_{\tilde{q}_{d,j}}, m_{\tilde{q}_{d,j}}, m_{\tilde{q}_{u,i}}]\\
&-&(D_{12}-D_{13}+2D_{22}+2D_{23} -4D_{26}) [p_1, p_2 - p_3, -p_2,
m_{\tilde{q}_{d,j}}, m_{\tilde{q}_{u,i}}, m_{\tilde{q}_{d,j}}, m_{\tilde{q}_{u,i}}]\\
&-& (D_{12}-D_{13}+2D_{22}+2D_{23} -4D_{26})[p_1, p_2 - p_3, -p_2,
m_{\tilde{q}_{u,i}}, m_{\tilde{q}_{d,j}}, m_{\tilde{q}_{u,i}}, m_{\tilde{q}_{d,j}}]\}\}
\hskip 7mm (A-5)
\end{eqnarray*}

$$f_{6}=f_{7}= \cdots=f_{12}=0
\eqno(A-6)$$

$$ A_{H^0} = \frac{i}{s - m_{H^0}^2 + i \Gamma_{H^0} m_{H^0}}, $$
$$ A_{h^0} = \frac{i}{s - m_{h^0}^2 + i \Gamma_{h^0} m_{h^0}}. \eqno(A-7) $$
The coupling constants are obtained by converting Feynman rules for
$X\tilde{q}\tilde{q}$(where X is a one- or two-particle state) from
the $\tilde{q}_{L} - \tilde{q}_{R}$ basis, their Feynman rules can be found
in Ref.\cite{haber1}, to the $\tilde{q}_{1} - \tilde{q}_{2}$ basis.
For the converting formulae one can refer to reference\cite{haber}.

In above expressions we adopted the definitions of one-loop integral
functions in reference \cite{s13} and defined $d=4-2 \epsilon$.
The arguments of two-point, three-point and four-point integral functions
are written at the end of formulae in brackets. The numerical calculation
of the vector and tensor loop integral functions can be traced back to four
scalar loop integrals $A_{0}$, $B_{0}$, $C_{0}$, $D_{0}$ as shown in the
reference\cite{s14}.

\vskip 20mm
{\Large{\bf Figure captions}}
\vskip 5mm
\noindent
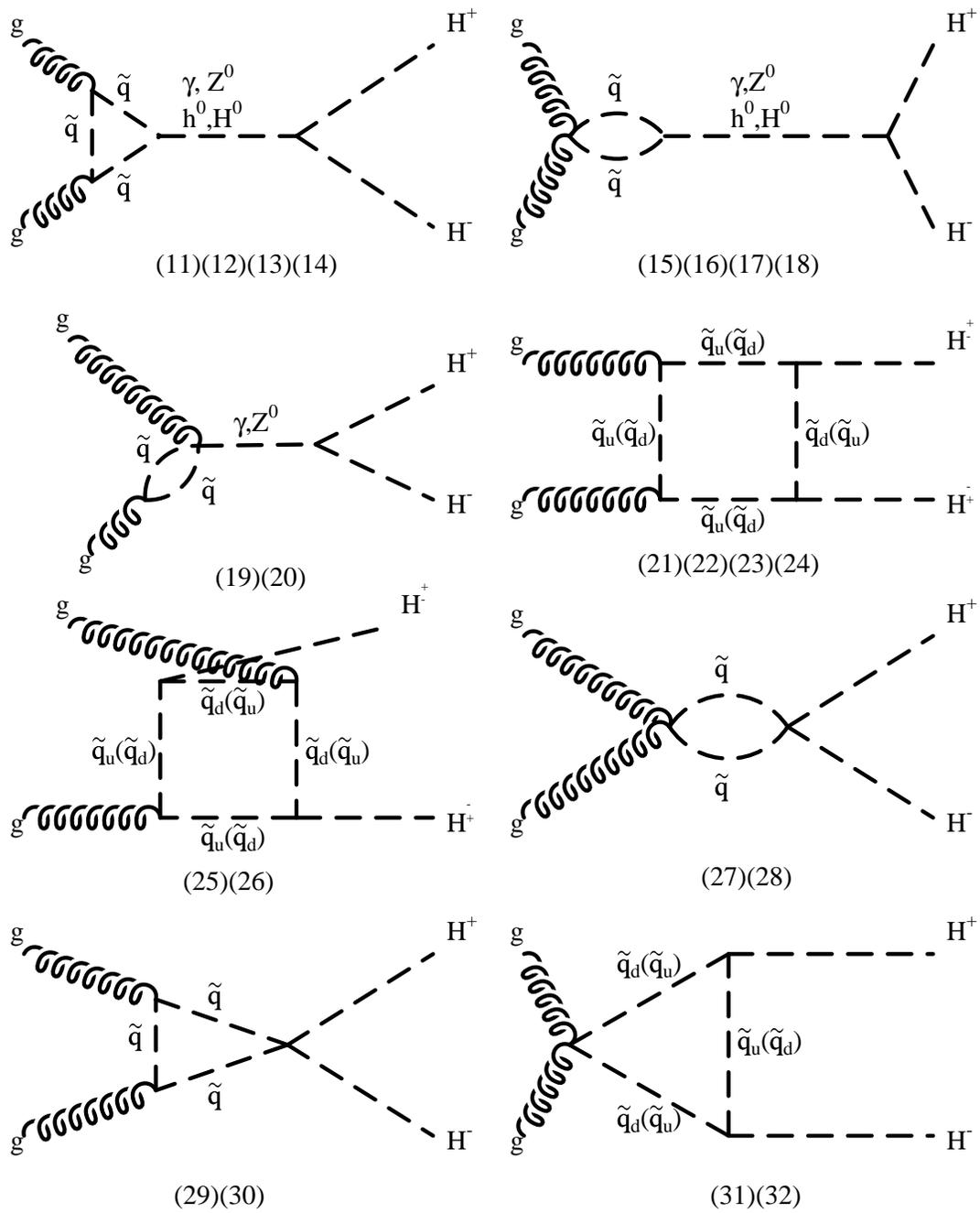
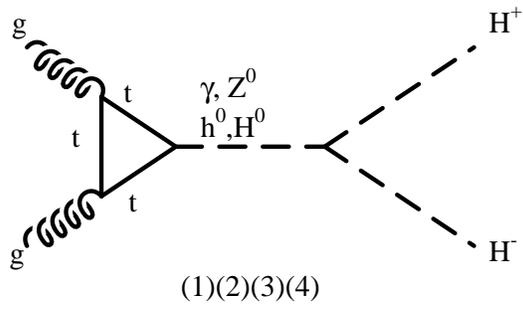
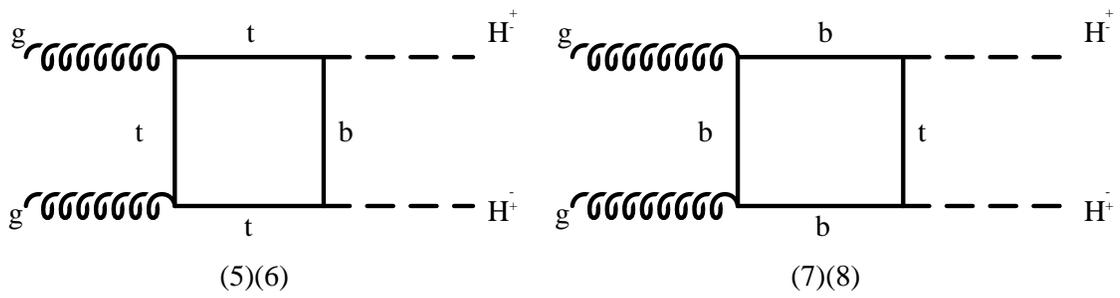
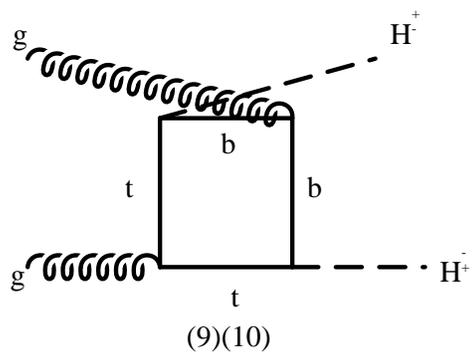
{\bf Fig.1} The Feynman diagrams of subprocess $gg \rightarrow H^{+} H^{-}$.
Figures ($1 \sim 10$) are so called THDM diagrams. Figures ($11 \sim 32$)
are the diagrams having one squark-loop.
($1 \sim 4$, $11\sim 20$, $27\sim 32$) are s-channel diagrams. The others
are box diagrams which include t- and u-channels.
\vskip 3mm
\noindent
{\bf Fig.2(1)} Total cross sections of the subprocess $gg \rightarrow
H^{+} H^{-}$ as the functions of the $\sqrt{\hat{s}}$ with
$m_{H^{\pm}}=150~GeV$ in the MSSM. The full-line is for $\tan{\beta}=1.5$.
The dotted-line is for $\tan{\beta}=30$ .
\vskip 3mm
\noindent
{\bf Fig.2(2)} Total cross sections of the subprocess $gg \rightarrow
H^{+} H^{-}$ as the functions of the $\sqrt{\hat{s}}$ with
$m_{H^{\pm}}=300~GeV$ in the MSSM. The full-line is for $\tan{\beta}=1.5$.
The dotted-line is for $\tan{\beta}=30$.
\vskip 3mm
\noindent
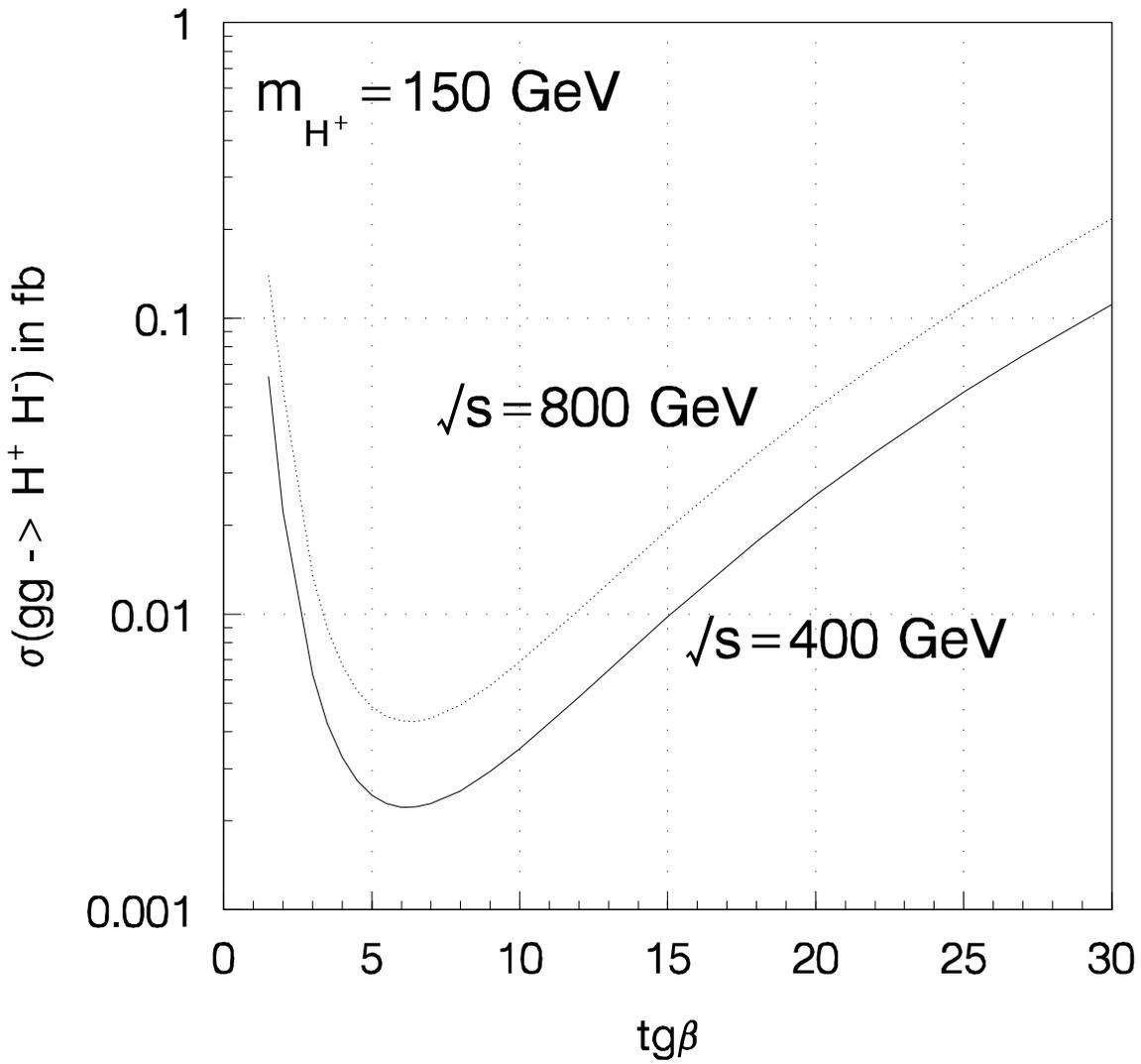
{\bf Fig.3} Total cross sections of the subprocess $gg \rightarrow
H^{+} H^{-}$ as the functions of the $\tan{\beta}$ with
$m_{H^{\pm}}=150~GeV$ in the MSSM.
The full-line and dotted-line correspond to $\sqrt{\hat{s}}=400~GeV$ and
$\sqrt{\hat{s}}=800~GeV$ respectively.
\vskip 3mm
\noindent
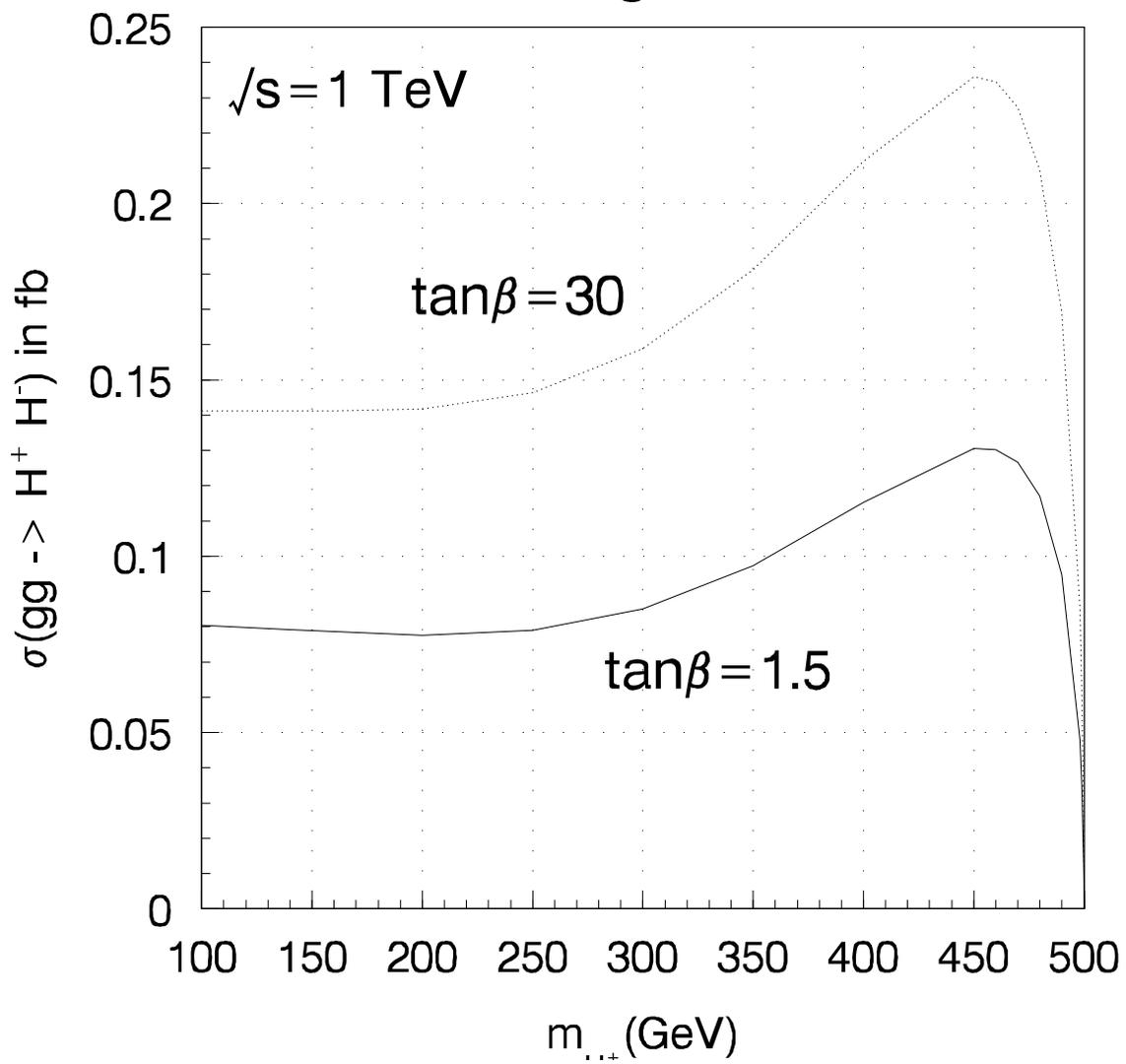
{\bf Fig.4} Total cross sections of the subprocess $gg \rightarrow
H^{+} H^{-}$ as the functions of the $m_{H^{\pm}}$ with
$\sqrt{\hat{s}}=1~TeV$ in MSSM. The full-line is for $\tan{\beta}=1.5$.
The dotted-line is for $\tan{\beta}=30$.
\vskip 3mm
\noindent
{\bf Fig.5(1)} Total cross sections of the process $pp \rightarrow gg
\rightarrow H^{+} H^{-} + X$ as the functions of the $\sqrt{s}$ with
$m_{H^{\pm}}=150~GeV$ at the LHC energy region. The full-line is for
$\tan{\beta}=1.5$ in the MSSM. The dashed-line is for $\tan{\beta}=1.5$
in the THDM. The dotted-line is for $\tan{\beta}=30$ in the MSSM. The
dashed-dotted-line is for $\tan{\beta}=30$ in the THDM.

\vskip 3mm
\noindent
{\bf Fig.5(2)} Total cross sections of the process $pp \rightarrow gg
\rightarrow H^{+} H^{-} + X$ as the functions of the $\sqrt{s}$ with
$m_{H^{\pm}}=300~GeV$ at the LHC energy region. The full-line is for
$\tan{\beta}=1.5$ in the MSSM. The dashed-line is for $\tan{\beta}=1.5$ in
the THDM. The dotted-line is for $\tan{\beta}=30$ in the MSSM. The
dashed-dotted-line is for $\tan{\beta}=30$ in the THDM.

\vskip 3mm
\end{large}

\newpage
\begin{thebibliography}{s25}
\bibitem{s1} S.L. Glashow, Nucl. Phys. 22(1961)579; S. Weinberg, Phys. Rev.
             Lett. 1(1967)1264; A. Salam, Proc. 8th Nobel Symposium Stockholm
             1968, ed. N. Svartholm(Almquist and Wiksells, Stockholm 1968)
             p.367; H.D. Politzer, Phys. Rep. 14(1974)129.
\bibitem{s2} P.W. Higgs, Phys. Lett 12(1964)132, Phys. Rev. Lett. 13
             (1964)508; Phys.Rev. 145(1966)1156; F.Englert and R.Brout,
             Phys. Rev. Lett. 13(1964)321; G.S. Guralnik, C.R.Hagen
             and T.W.B. Kibble, Phys. Rev. Lett. 13(1964)585; T.W.B. Kibble,
             Phys. Rev. 155(1967)1554.
\bibitem{s3} H. E. Haber, G. L. Kane, Phys. Rep. 117(1985) 75.
\bibitem{haber}J.F. Gunion and H.E. Haber, Nucl. Phys. {\bf B272},1(1986).
\bibitem{s4} T.Plehn, M.Spira and P.M. Zerwas, Nucl. Phys. {\bf B479},46(1996).
\bibitem{s5} D.Bowser-Chao, K.Cheung, and S.Thomas, Phys. Lett {\bf B315},
             399(1993); Ma Wen-Gan, Chong-Sheng Li and Hang Liang, Phys. Rev.
             {\bf D53},1304(1996),[E:1997, Phys. Rev. D56, 4420].
\bibitem{s6} Jiang Yi, Han Liang, Ma Wen-Gan, Yu Zeng-Hui and Han Meng,
             J. of Phys. {\bf G23},385(1997), [E:1997, J. Phys.
             G: Nucl. Phys. 23, 1151].
\bibitem{s7} R.Arnowitt, A.H. Chamseddine and P.Nath, Phys. Rev. Lett,
             49,970(1982); 50,232(1983); Phys. Lett. {\bf B121},33(1983);
             H.P.Nilles, Phys. Rep. 110,1(1984); J. Ellis, C.Kounnas and D.V.
             Nanopoulos, Nucl. Phys. {\bf B241},406(1984).
\bibitem{s8} Scott.S.D.Willenbrock, Phys. Rev. {\bf D 35},173(1987).
\bibitem{Krause} A. Krause, T Plehn, M. Spira and P.M.Zerwas, `Pair
         production of charged Higgs boson pairs in gluon-gluon collisions`,
         hep-ph/9707430.
\bibitem{s9} J. Ellis and S. Rudaz, Phys. Lett. {\bf B128},248(1983);
             F. Gunion and H.E. Haber, Nucl. Phys. {bf B272},1(1986).
\bibitem{s10}H.E. Haber, G.L. Kane, Phys. Rep. 117(1985)75.
\bibitem{s11}M. Gell-Mann, Phys. Rev. {\bf 125},1067(1962).
\bibitem{ss1}A.D. Martin, W.J. Stirling and R.G. Roberts, Phys. Lett.{\bf B354},
             155(1995).
\bibitem{s12}Particle Data Group, Phys. Rev. {\bf D 50}, No.3(1994);
             D. Schaile, CERN-PPE/94-162(11 October 1994).
\bibitem{s13}Bernd A. Kniehl, Phys. Rep. 240(1994)211.
\bibitem{s14}G. Passarino and M. Veltman, Nucl. Phys. {\bf B160}, 151(1979).
\bibitem{haber1} J.F. Gunion, H.E. Haber, G. Kane and S. Dawson, The Higgs
             Hunter's Guide, Addison Wesley, Reading, MA(1990);
\end{thebibliography}
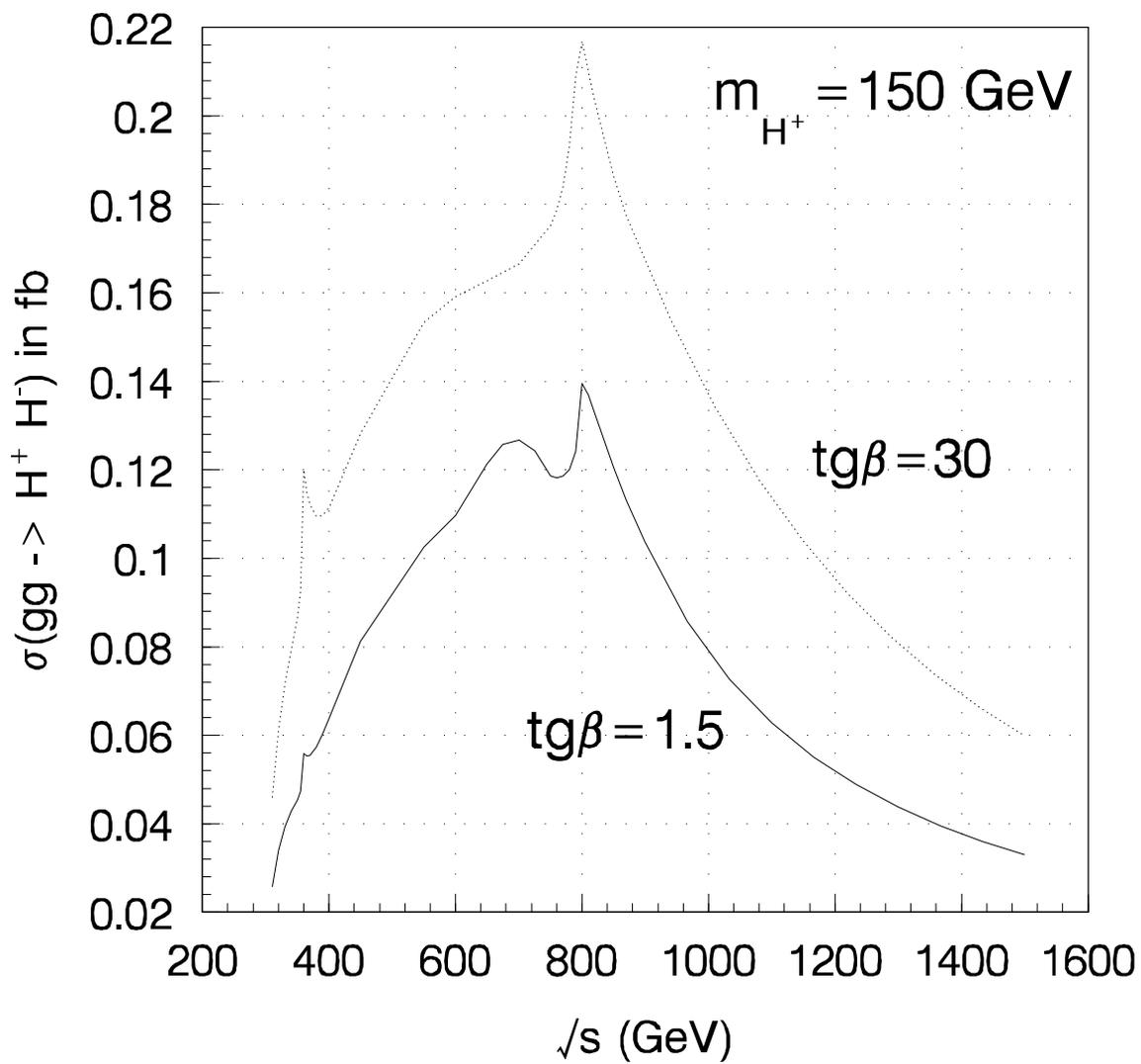
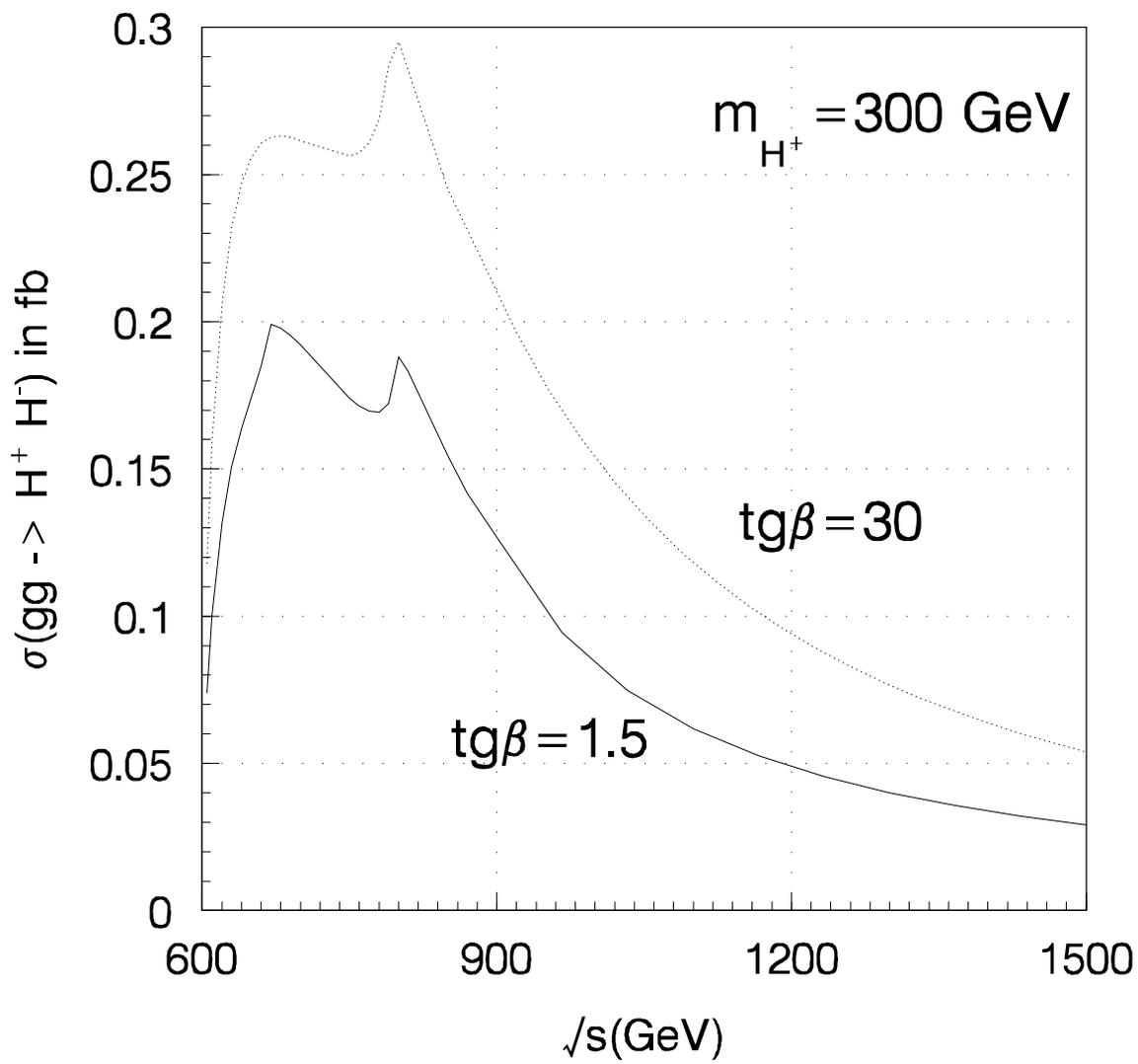
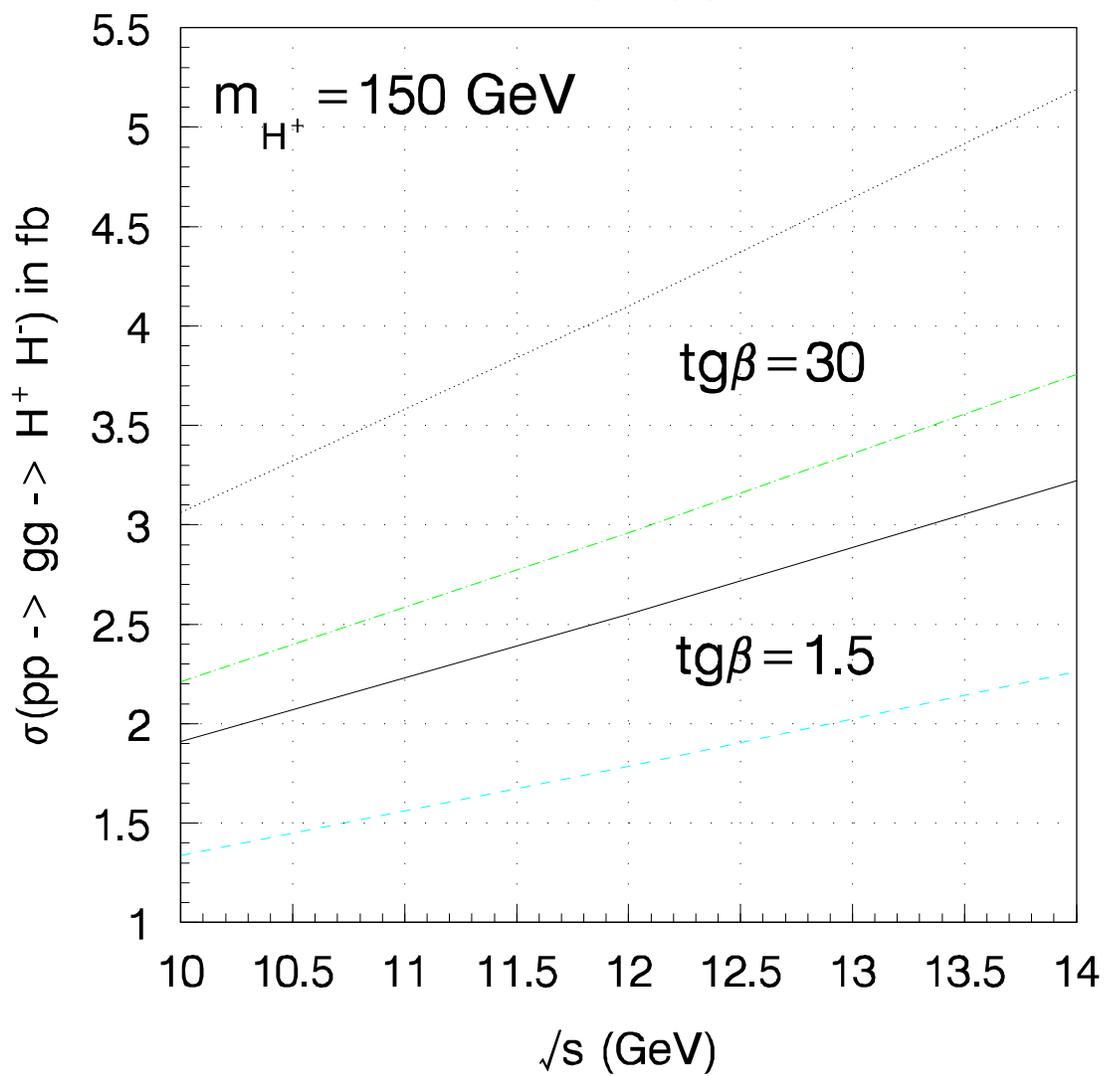
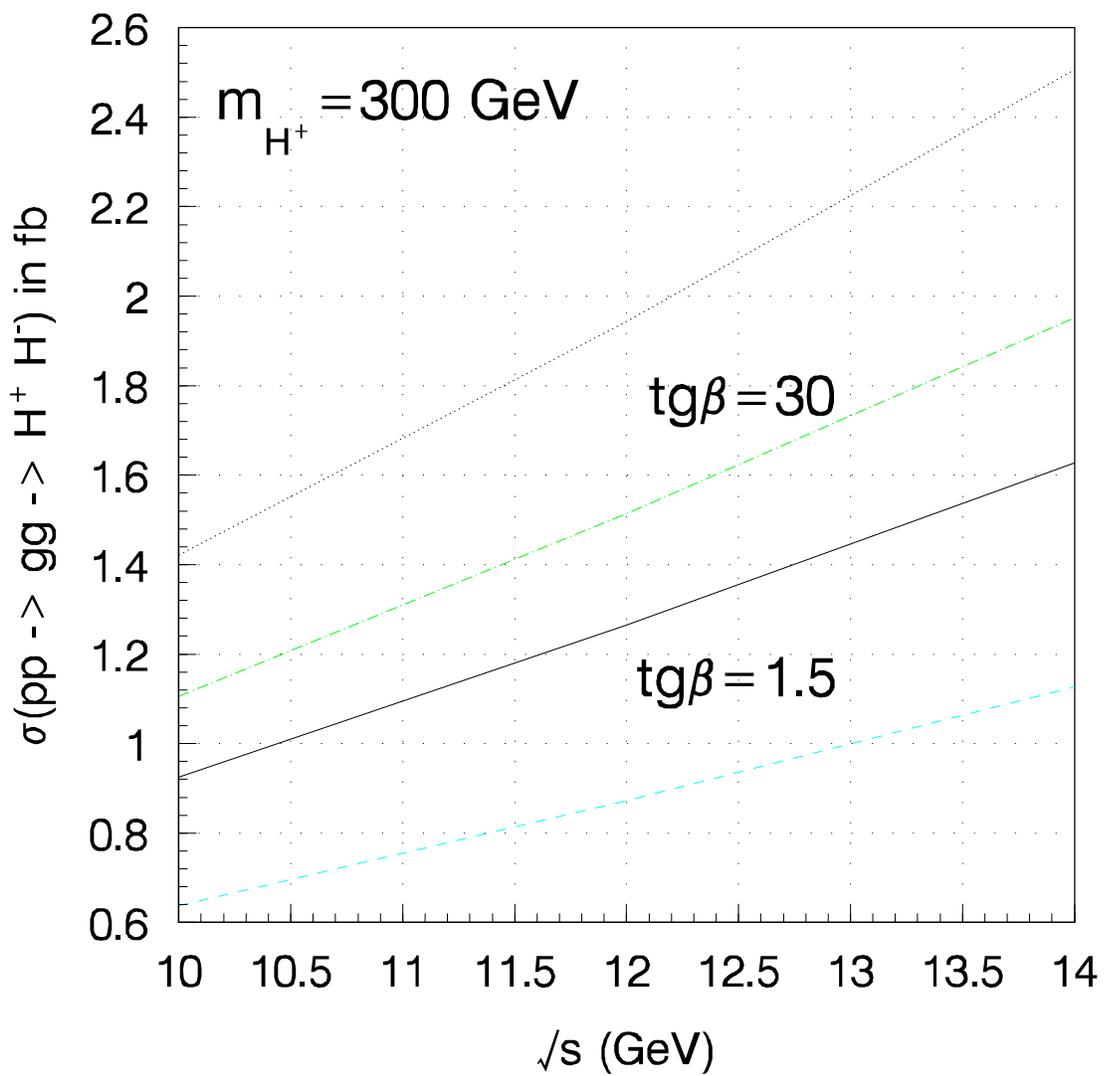
\end{document}